\documentclass[12pt]{iopart}
\pdfoutput=1
\usepackage{iopams}
 
\expandafter\let\csname equation*\endcsname\relax
\expandafter\let\csname endequation*\endcsname\relax
\usepackage{amsmath,amssymb,mathtools}

\usepackage{graphicx,color}
\usepackage{enumerate}

\usepackage[colorlinks,linkcolor=blue,urlcolor=blue,citecolor=blue]{hyperref}

 \newcommand{\erefss}[1]{\eqref{#1}}%
 \newcommand{\erefs}[1]{\eqref{#1}}%
 \newcommand{\ffref}[1]{Figure~\ref{#1}} %
 \newcommand{\srefss}[1]{Secs.~\ref{#1}}%
 \newcommand{\srefs}[1]{~\ref{#1}}%
 \newcommand{\ssref}[1]{Section~\ref{#1}}%

\newcommand\be{\begin{equation}} 
\newcommand\ee{\end{equation}} 
\begin{document}

\title{Stochastic resetting in underdamped Brownian motion}

\author{Deepak Gupta}  
\address{Raman Research Institute,  Bangalore - 560080, India}

\date{\today}

\begin{abstract}
We consider a single Brownian particle in one dimension in a medium at a constant temperature in the underdamped regime. We stochastically reset the position of the Brownian particle to a fixed point in the space with a constant rate $r$ whereas its velocity evolves irrespective of the position of the particle and stochastic resetting mechanism. The nonequilibrium  steady state of the position distribution is studied for this model system. Further, we study the distribution of the position of the particle in the finite time and the approach to the nonequilibrium steady state distribution with time. Numerical simulations are done to verify the analytical results.
\end{abstract}


\maketitle

\noindent\rule{\hsize}{2pt}
\tableofcontents
\noindent\rule{\hsize}{2pt}
\markboth{Stochastic resetting in underdamped Brownian motion}{}

\section{Introduction}
\label{intro}
Search strategies have gained much attention over the past few years. These can be seen in many branches of science. For example, in ecology, one is interested in understanding the nature of animal foraging for food in the forest \cite{book-search,search-strat,Catalan}. In the context of microscopic organism, the \emph{E. coli} does run and tumble motion to reach the higher food concentration region \cite{bergecoli}. In computer science, optimization of the computer search algorithms for hard combinatorial problems is an efficient strategy  \cite{Montanari}. In biology, transcription factor (a protein) utilizes both three and one dimension diffusion processes (the \emph{facilitated diffusion}) along the DNA strand to find their DNA promoter site \cite{david}. In general, these strategies involve two type of processes: an exploration of the local area and long jumps to a prescribed domain (i.e., switching between states). Within this context, several physical examples one can see in a large number of situations. In daily routine, searching objects such as a misplaced object at home, a thief in the city, a friend on a railway platform, etc., are common examples. The searcher tries to find the target object for a while, upon the unsuccessful attempt, he/she returns to the initial location and restarts the process of searching. Such a manner of exploring a given area is referred to as \emph{stochastic resetting}. There are other examples which involve such mechanism. For example, the persistence shown by sub-population of bacteria by switching between different states to save the population from the extinction in the changing environment \cite{Kussell1807}. Similarly, microbes switch between two phenotype states with different growth rates to survive in the sudden catastrophic events in the environment whose rate depends on the microbial population \cite{Visco}. Such stochastic resetting mechanism is introduced for a walker having long-range memory \cite{Boyer}. Resetting is also used to understand the biological reactions in the presence of enzymes \cite{Reuveni4391}.  Similar studies one can see in Refs. \cite{Gated,Lomholt,Loverdo,Gleb,berg}. 

Consider a Brownian motion in one dimension. The variance of the position of the Brownian particle grows with the observation time, i.e.,  $\langle [x(t)-\langle x(t) \rangle]^2\rangle \sim t$. Therefore, the position distribution (which is Gaussian) never reaches a steady state. Moreover, the mean first passage time of the Brownian particle to reach an absorber at $x$ starting from $x_0$ is infinite \cite{redner2001}. On introducing the stochastic resetting mechanism where in addition to the diffusion of a particle one resets the position of the particle to a desired location with a constant rate $r$, in the long time limit, the position distribution reaches a nonequilibrium  steady state which has non-Gaussian fluctuations. Moreover, the mean first passage time for a particle following such a dynamics to find an absorber achieves a finite value \cite{Satya-1}. Both of these results are quite remarkable. The approach to steady state distribution with time for a Brownian particle under the resetting mechanism is studied in Ref. \cite{sanjib-relaxation}. An extension of results in \cite{Satya-1} with space-dependent resetting, resetting at random position, and spatial distribution of the absorbing target is presented in Ref. \cite{satya-optimal}. Later, a generalization of \cite{Satya-1} for an arbitrary dimension is given in Ref. \cite{Evans-satya-dim}. Recently, there has been a lot of studies made to understand the diffusion with stochastic resetting with a number of different settings. For example,  diffusion with stochastic resetting in bounded domain \cite{christou}, and inside a circle \cite{Abhinava-circle} are investigated. Majumdar et al. \cite{reset-max-pos} studied a model of a one dimension lattice random walk where the walker resets at the maximum of the already visited positions with a probability $p$ otherwise it does a symmetric random walk. Evans and Majumdar \cite{refractory} computed the steady state distribution for a particle which undergoes the resetting mechanism and stays at the resetting location during a random refractory period.
Montero et al. \cite{Montero} studied the continuous time random walks in the presence of drift and stochastic resetting. Kusmierz et al. \cite{levy-reset} investigated the mean first passage time for a searcher whose jumps are drawn from arbitrary distribution and reset to a given location with probability $p\in[0,1)$ to find a target. Moreover, when the jump distribution has heavy-tailed L\'evy distribution, they found a global minimum for the mean first passage time  in the parameter space. Gupta et al. \cite{shamik-interface} studied a fluctuating interface model with random resetting to its initial profile. They computed the steady state distribution of the height profile. Later, this model is generalized for a resetting time drawn from power-law distribution \cite{KPZ-power-law}. 
Pal \cite{Arnab-1} studied the steady state distribution of the position of a diffusing particle with resetting in a potential landscape. In the presence of time-dependent resetting, Pal et al. \cite{pal-time-dep} investigated the steady state distribution of the position of the particle, relaxation to the steady state, and mean first passage time. Nagar et al. \cite{shamik-apoorva} studied a model of a diffusing Brownian particle resets to the origin at a time drawn from a power-law distribution. Bhat et al. \cite{Uttam} introduced two types of resetting mechanism: Poisson and deterministic resetting to $N$ number of diffusing particle in one, two, and three dimensions and studied the mean first passage time.  Falcao et al. \cite{falcao-inter} studied two interacting particles in one dimension with a bias towards each other with a condition that they get reset to their initial location once they are about to colloid each other. In this model, they studied the steady state distribution and relaxation of the time-dependent distribution towards it. Path integral formalism for stochastic resetting recently introduced by Rold\'an et al. \cite{roldan-path-int}. Rose et al. \cite{Rose} studied the spectral properties of classical and quantum Markov processes which resets at random times. Similarly, one can also find several studies where stochastic resetting mechanism is imposed in Refs. \cite{under-restart,Manrubia,Shlomi-PRL,Rolden-RNA,evans-satya-kirone,run-tumble,PRB,vvpal}.   Recently, some investigations are also devoted in understanding the large deviation function of the observable of the Markov processes with stochastic resetting, and connection of stochastic thermodynamics \cite{Sekimoto,Seifert-1,Seifert-2} and resetting \cite{Fuchs}. Within this context, Pal et al. \cite{IFT-arnab} studied the Hatano-Sasa relation for the system undergoes steady state transitions and integral fluctuation theorem in the presence of the stochastic resetting. 

In this paper, we consider a single Brownian particle in one dimension in the underdamped regime. The state of the particle is described by the position and velocity variables $(x,v)$ at a time $t$. The position of the particle resets to the initial location of the particle with a constant rate $r$ whereas the velocity of the particle does not get affected by the resetting mechanism. Therefore, the velocity variable of the particle enjoys the Gibbs-Boltzmann distribution at a long time. We study the steady state distribution of the position variable analytically using the \emph{renewal process}. Moreover, we analytically compute the large but finite time distribution of the position variable and study the relaxation to the steady state distribution. Numerical simulations are done to compare the analytical predictions.

The remaining paper is structured as follows. In \sref{model-1}, we discuss the model of a single Brownian particle in one dimension whose position is reset to the initial location of the particle while the velocity variable undergoes the evolution independent of the position and stochastic resetting mechanism. \ssref{sec-ss} contains the steady state distribution. In \srefss{prop-sec-1} and \srefs{prop-sec-2}, we compute the propagators. We obtain the tail of the steady state distribution in \sref{sec-asym}. In \sref{sec-rlx}, we compute the analytic probability density function for the position of the Brownian particle in the large but finite time. Finally, we summarized our paper in \sref{summ}.

\section{Model}
\label{model-1}
Consider a single Brownian particle of mass $m$ immersed in a heat bath at a temperature $T$. For simplicity, we consider the motion along one dimension. The evolution of the particle is given by the underdamped Langevin equations:
\begin{align}
&\dot x=v(t)\label{pos-1},\\
&m\dot v=-\gamma v(t)+\eta(t) \label{vel-1},
\end{align}
where $x(t)$ and $v(t)$ are the position and velocity of the Brownian particle at time $t$. In the above equations, the dot indicates the total time derivative, $\gamma$ is the dissipation constant, and $\eta(t)$ is the Gaussian thermal white noise from the heat bath with properties $\langle \eta(t) \rangle=0$, and $\langle \eta(t_1)\eta(t_2) \rangle=2D\gamma^2 \delta (t_1-t_2)$, where $D=k_B T/\gamma$ is the Diffusion constant, and $k_B$ is the Boltzmann's constant.

\begin{figure}
\begin{center}
    \includegraphics[width=18cm]{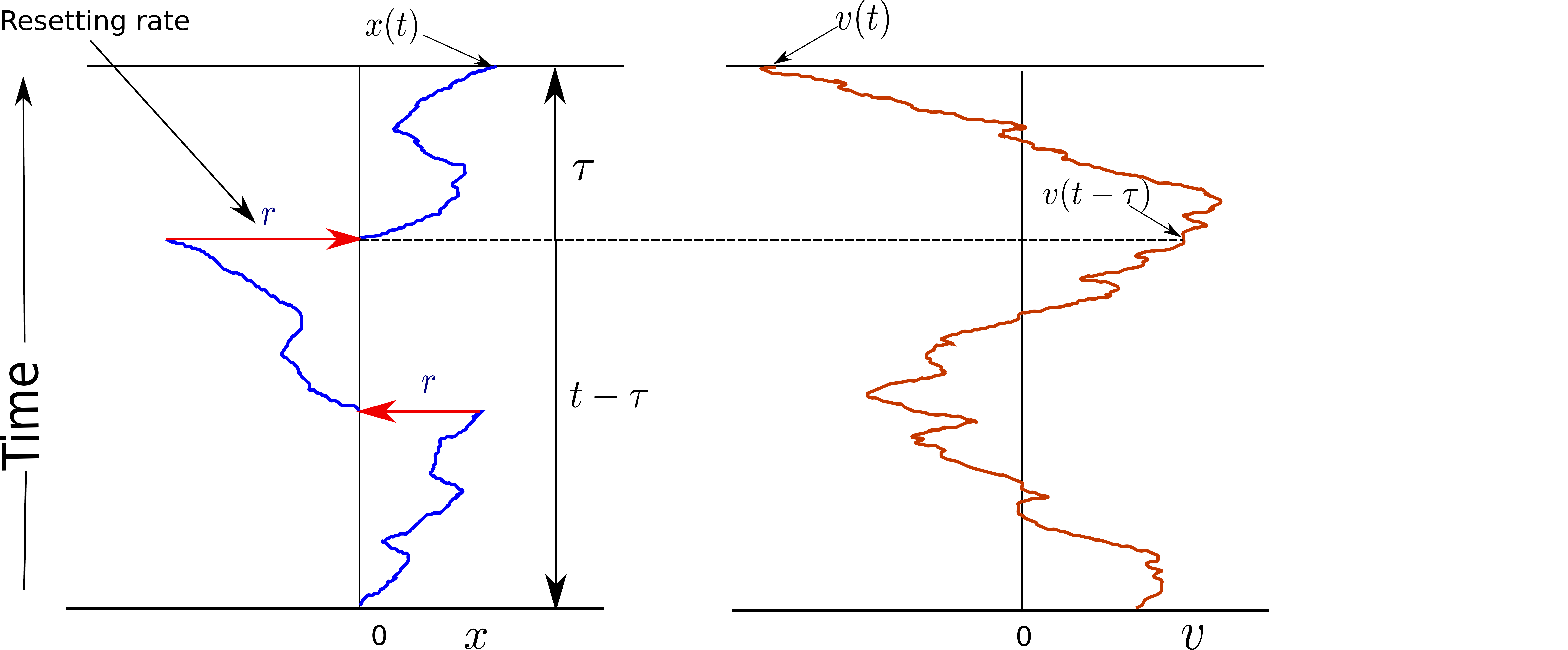}
\end{center}
   \caption{\label{setup-1} A realization of the position $x$ and the velocity $v$ of the Brownian particle with respect to time is shown. The velocity of the particle evolves according to \eref{vel-1} whereas the position of the particle evolves with \eref{pos-1} and stochastically reset to the origin of the $x$-axis with a constant rate $r$ [see \eref{reset-1}]. Here, $t-\tau$ is the time where the last reset occurs, and the remaining time [$t-\tau,t$] the Brownian particle evolves as specified in \eref{pos-1}. }
\end{figure}

Suppose we reset the position of the Brownian particle to the initial position $x(0)=0$ with a rate $r$ whereas the velocity of the particle evolves as \eref{vel-1}.  Therefore, in a time increment $\Delta t$, we see that
\begin{align}
x(t+\Delta t)=\begin{cases}
x(0) \quad\quad\quad\quad\quad \text{with probability} \quad\quad r\Delta t,\\
x(t)+v(t) \Delta t\quad \text{with probability} \quad\quad (1-r\Delta t),
\label{reset-1}
\end{cases}
\end{align}
whereas 
\begin{align}
v(t+\Delta t)=(1-\Delta t/\tau_\gamma)v(t)+\sqrt{2D\gamma^2\Delta t}\ \tilde\eta_{\Delta t}(t)/m,
\label{v-reset}
\end{align}
where $\tau_\gamma=m/\gamma$ is the relaxation time, and $\tilde\eta_{\Delta t}(t)$ is a Gaussian random variable with mean zero and variance one at each time $t$, i.e.,   $\tilde\eta_{\Delta t}(t)=\mathcal{N}(0,1)$. The schematic diagram of the above process is shown in \fref{setup-1}.

Since the velocity variable is linear in the Gaussian thermal white noise [see \eref{vel-1}], the probability density function $p(v,t|v_0)$ of the velocity $v$ of the Brownian particle at time $t$ with initial velocity $v_0$  has the Gaussian distribution:
\begin{align}
p(v,t|v_0)=\frac{1}{\sqrt{2 \pi \Delta^2(t)}}\exp\bigg[-\frac{(v-v_0e^{-t/\tau_\gamma})^2}{2 \Delta^2(t)}\bigg],
\end{align}
where $\Delta^2(t):=\langle [v(t)-\langle v(t)\rangle ]^2\rangle= \frac{D}{\tau_\gamma}(1-e^{-t/\tau_\gamma})$.

Thus, in the large time limit ($t\gg \tau_\gamma$), the velocity variable achieves the steady state distribution:
\begin{align}
p_{ss}(v)=\sqrt{\frac{ \tau_\gamma}{2\pi D}} \exp\bigg[{-\frac{\tau_\gamma v^2}{2D}}\bigg].
\label{ss-v0}
\end{align}
On the other hand, the position variable undergoes the resetting mechanism as described in \eref{reset-1} and \fref{setup-1}. Therefore, the probability density function of it is not expected to be Gaussian. Our main aim in the paper is to understand the distribution of the position variable in the presence of stochastic resetting.


\section{Steady state distribution: $P_r^{ss}(x)$}
\label{sec-ss}
To obtain the probability density function for $x$, we first write the Fokker-Planck equation for the joint conditional probability density function $\rho_r(x,v,t|x_0,v_0)$ as
\begin{align}
\dfrac{\partial \rho_r(x,v,t|x_0,v_0)}{\partial t}=& \bigg[-v\dfrac{\partial}{\partial x}+\dfrac{1}{\tau_\gamma}\dfrac{\partial}{\partial v}v+\dfrac{D}{\tau_\gamma^2}\dfrac{\partial^2}{\partial v^2}\bigg]\rho_r(x,v,t|x_0,v_0)\nonumber\\&-r\rho_r(x,v,t|x_0,v_0)+r\delta(x-x_0)\int_{-\infty}^{+\infty}dx'\rho_r(x',v,t|x_0,v_0),
\label{FP-eqn}
\end{align}
where the first three terms on the right hand side correspond to \erefss{pos-1} and \erefs{vel-1} (without the resetting mechanism), fourth term represents the negative flux from each point $(x,v)$, and the fifth term is the gain term indicating the positive flux into the point $x=x_0$ (i.e., resetting location and initial position are same). Notice that in our case, we do not reset the velocity variable. The above Fokker-Planck equation is subjected to the initial condition $\rho_r(x,v,t=0|x_0,v_0)=\delta(x-x_0)\delta(v-v_0)$.

The probability density function for $x$ at time $t$ is obtained by integrating the joint condition distribution, i.e., the solution of the Fokker-Planck equation \eref{FP-eqn}], as
\begin{equation*}
P_r(x,t)=\int_{-\infty}^{+\infty} dv\ \int_{-\infty}^{+\infty} dv_0\ \int_{-\infty}^{+\infty} dx_0\ \rho_r(x,v,t|x_0,v_0) \rho(x_0,v_0),
\end{equation*}
where $\rho(x_0,v_0)$ is the initial distribution of $x_0$ and $v_0$.

In the large time limit ($t\to\infty$), one can obtain the steady state distribution for the position of the Brownian particle using $P_r(x,t)$.

We can obtain the steady state distribution using the \emph{renewal} process \cite{Satya-1,sanjib-relaxation}. In the renewal picture, the probability density function of $x$ at time $t$ when the particle is stochastically reset to  the initial position $x_0$ with a constant rate $r$ is given by
\begin{align}
P_r(x,t|x_0)=e^{-rt}P^{\mathrm{(I)}}_0(x,t|x_0)+r \int_0^t d\tau e^{-r\tau}P^{\mathrm{(II)}}_0(x,\tau|x_0),
\label{renewal-eqn}
\end{align}
where on the right hand side, the first term is the contribution from those events when there has been no reset between time $t=0$ to $t=t$ with the probability of no reset $e^{-rt}$ [i.e., the particle moves from $x_0$ to $x$ following \erefss{pos-1} and \erefs{vel-1}], and the second term is due to those events when there have been multiple resets happened with the last reset occurred at time $t-\tau$ and subsequently the particle follows the dynamics \erefss{pos-1} and \erefs{vel-1} for a time duration $\tau$ (the probability of which is $r e^{-r \tau}$) and reaches at $x$ (see \fref{setup-1}). In the above equation, $P^{\mathrm{(I,II)}}_0(x,u|x_0)$ is the free propagator at time $u$ obtained from \erefss{pos-1} and \erefs{vel-1} (see \fref{setup-1}).

In the large time limit ($t\to\infty$), the dominant contribution to the steady state distribution of the position of the Brownian particle is
\begin{align}
P_r^{ss}(x)=r \int_0^\infty d\tau e^{-r\tau}P^{\mathrm{(II)}}_0(x,\tau|x_0).
\label{ss-reset}
\end{align}

\subsection{The propagator $P^{\mathrm{(I)}}_0(x,t|x_0)$}
\label{prop-sec-1}
The contributions to the propagator $P^{\mathrm{(I)}}_0(x,t|x_0)$ come from those trajectories which start from $x_0$ and reach $x$ without resetting up to time $t$. We can obtain it as follows. For convenience, we choose the initial location of the particle at the origin of the $x$-axis. Therefore, the position of the particle at time $t$ is
\begin{equation}
x(t)=\int_{0}^{t}dt_1\ v(t_1),
\label{x-reset}
\end{equation}
where $v(t_1)$ is the velocity of the Brownian particle at time $t=t_1$:
\begin{align}
v(t_1)=v_0 e^{-t_1/\tau_\gamma}+\dfrac{1}{m} \int_0^{t_1} dt_1^\prime e^{-(t_1-t_1^\prime)/\tau_\gamma}. 
\label{v-evol}
\end{align}
From equation \eref{x-reset}, we can see that $x$ is linear in the Gaussian variable $v$. Therefore, the distribution of $x$ would be Gaussian, and the mean and the variance of $x$ are sufficient to write $P^{\mathrm{(I)}}(x,t|x_0=0,v_0,0)$ which are given as following
\begin{align}
\langle x(t)\rangle=&\int_{0}^{t}dt_1\ \langle v(t_1)\rangle,\\
\langle [x(t)-\langle x(t)\rangle]\rangle^2=&\int_{0}^{t}dt_1\int_{0}^{t}dt_2\ [\langle v(t_1)v(t_2)\rangle-\langle v(t_1)\langle v(t_2)]\label{var-x},
\end{align}
where the angular brackets show the average over the Gaussian thermal white noise. Therefore, we see that
\begin{align}
&\langle v(t_1)\rangle=v_0 e^{-t_1/\tau_\gamma},\label{eq-1}\\
&\langle v(t_1)v(t_2)\rangle=v_0^2 e^{-(t_1+t_2)/\tau_\gamma}+\dfrac{D}{\tau_\gamma}(e^{-|t_1-t_2|/\tau_\gamma}-e^{-(t_1+t_2)/\tau_\gamma})\label{eq-2}.
\end{align}
Using \erefss{eq-1} and \erefs{eq-2}, we find that 
\begin{align}
\langle x(t)\rangle=&v_0\tau_\gamma(1-e^{-t/\tau_\gamma}),\\
\Sigma_{\mathrm{(I)}}^2:=\langle [x(t)-\langle x(t)\rangle]\rangle^2=& D\tau_\gamma(2t/\tau_\gamma+4 e^{-t/\tau_\gamma}-e^{-2t/\tau_\gamma}-3),
\end{align}
Therefore, the probability density function of position $x$ starting from $x_0=0$ with the initial velocity $v_0$ is given as
\begin{equation}
P^{\mathrm{(I)}}(x,t|x_0=0,v_0,0)=\dfrac{1}{\sqrt{2 \pi \Sigma_{\mathrm{(I)}}^2}}\exp\bigg[-\dfrac{(x-\langle x(t) \rangle)^2}{2\Sigma_{\mathrm{(I)}}^2}\bigg].
\end{equation}
In our paper, we choose the initial velocity $v_0$ from the steady state distribution given by $p_{ss}(v_0)$ in \eref{ss-v0}. Therefore, 
\begin{align}
P^{\mathrm{(I)}}_0(x,t|x_0=0)&=\int_{-\infty}^{+\infty}dv_0\ P^{\mathrm{(I)}}(x,t|x_0=0,v_0,0) p_{ss}(v_0),\nonumber\\
 &=\dfrac{1}{\sqrt{2\pi \sigma^2(t)}}\exp\bigg[-\dfrac{x^2}{2\sigma^2(t)}\bigg],
\label{prop-1}
\end{align}
where 
\begin{equation}
\sigma^2(t)=2 D \tau_\gamma (t/\tau_\gamma+e^{-t/\tau_\gamma}-1).
\label{sigma}
\end{equation}

\subsection{The propagator $P^{\mathrm{(II)}}_0(x,\tau|x_0)$}
\label{prop-sec-2}
In this case, the contributions to the propagator $P^{\mathrm{(II)}}_0(x,\tau|x_0)$ are due to those trajectories which start from $x(t-\tau)=0$ (the last reset position) and reach $x$ (see \fref{setup-1}).  Therefore,
\begin{equation}
x(t)=\int_{t-\tau}^{t}dt_1\ v(t_1),
\end{equation}
where $v(t_1)$ is the velocity of the Brownian particle at time $t=t_1$ given in \eref{v-evol}.

The mean and the variance of $x$ are given as 
\begin{align}
\langle x(t)\rangle=&\int_{t-\tau}^{t}dt_1\ \langle v(t_1)\rangle,\\
\langle [x(t)-\langle x(t)\rangle]\rangle^2=&\int_{t-\tau}^{t}dt_1\int_{t-\tau}^{t}dt_2\ [\langle v(t_1)v(t_2)\rangle-\langle v(t_1)\langle v(t_2)],
\end{align}
Using \erefss{eq-1} and \erefs{eq-2}, we find that 
\begin{align}
\langle x(t)\rangle=&v_0\tau_\gamma e^{-t/\tau_\gamma}(e^{\tau/\tau_\gamma}-1),\\
\Sigma_{\mathrm{(II)}}^2:=\langle [x(t)-\langle x(t)\rangle]\rangle^2=& 2D\tau_\gamma(\tau/\tau_\gamma+e^{-\tau/\tau_\gamma}-1)-D\tau_\gamma e^{-2t/\tau_\gamma}(e^{\tau/\tau_\gamma}-1)^2,
\end{align}
Therefore, the probability density function of the position $x$ starting from $x(t-\tau)=0$ with the initial velocity $v_0$ is
\begin{equation}
P^{\mathrm{(II)}}(x,t|0,v_0,t-\tau)=\dfrac{1}{\sqrt{2 \pi \Sigma_{\mathrm{(II)}}^2}}\exp\bigg[-\dfrac{(x-\langle x(t) \rangle)^2}{2\Sigma_{\mathrm{(II)}}^2}\bigg].
\end{equation}
Averaging over the initial velocity $v_0$ with respect to the steady state distribution $p_{ss}(v_0)$ yields 
\begin{align}
P^{\mathrm{(II)}}_0(x,\tau|x_0=0)&=\int_{-\infty}^{+\infty}dv_0\ P^{\mathrm{(II)}}(x,t|0,v_0,t-\tau) p_{ss}(v_0),\nonumber\\
 &=\dfrac{1}{\sqrt{2\pi \sigma^2(\tau)}}\exp\bigg[-\dfrac{x^2}{2\sigma^2(\tau)}\bigg],
 \label{prop-2}
\end{align}

\subsection{Steady state distribution $P_r^{ss}(x)$ at large-$x$}
\label{sec-asym}
The steady state distribution $P_r^{ss}(x)$ is given in the integral form in \eref{ss-reset}. The analytical expression of $P_r^{ss}(x)$ is difficult to obtain. Nevertheless, one can compute it numerically. In the following, we discuss the large-$x$ behaviour of the $P_r^{ss}(x)$.

The Fourier transform of \eref{ss-reset} is given as
\begin{align}
\tilde{P}_r^{ss}(k)&=\int_{-\infty}^\infty dx\  e^{ikx}P_r^{ss}(x) , \nonumber\\
&=r\int_0^\infty d\tau\ e^{-r\tau}e^{-\frac{k^2\sigma^2}{2}},
\end{align}  
where $k$ is the conjugate variable with respect to $x$. For convenience, in the rest of the paper, we write $\sigma=\sigma(\tau)$.

Integration by parts of the above equation yields
\begin{align}
\tilde{P}_r^{ss}(k)=\dfrac{\alpha^2}{k^2+\alpha^2}\bigg[1+e^{D\tau_\gamma k^2}\int_0^\infty d\tau\ e^{-\tau D(k^2+\alpha^2)}\dfrac{d}{d\tau}\big(e^{-D\tau_\gamma k^2 e^{-\tau/\tau_\gamma}}\big)\bigg].
\label{F-pk}
\end{align}
The steady state distribution $P_r^{ss}(x)$ is obtained by inverting the above equation using inverse Fourier transform:
\begin{align}
P_r^{ss}(x)=\int_{-\infty}^{+\infty}\dfrac{dk}{2\pi}\dfrac{\alpha^2e^{-ikx}}{k^2+\alpha^2}\bigg[1+e^{D\tau_\gamma k^2}\int_0^\infty d\tau\ e^{-\tau D(k^2+\alpha^2)}\dfrac{d}{d\tau}\big(e^{-D\tau_\gamma k^2 e^{-\tau/\tau_\gamma}}\big)\bigg].
\label{inv-f}
\end{align}
The large-$x$ behaviour of the steady state distribution $P_r^{ss}(x)$ can be computed from the above equation as follows. In the above equation, there are two poles at $k=\pm i\alpha$ in the complex $k$-plane, where $\alpha=\sqrt{r/D}$ is the inverse length corresponding to the typical distance travelled by the particle between resets. Using Cauchy's residue theorem, one gets
\begin{equation}
P_r^{ss}(x) \to \dfrac{\alpha}{2}e^{-\alpha^2D\tau_\gamma}e^{-\alpha |x|}\quad\quad \text{as}\quad\quad |x|\to\infty.
\label{asym}
\end{equation}
The above result is the extension of the result given in \cite{Satya-1} for the overdamped case. From \eref{inv-f}, the steady state distribution of the position of a diffusing Brownian particle and stochastically  resetting  at the origin with a constant rate $r$ in the overdamped limit ($\tau_\gamma\to 0$) is obtained as $P_r^{ss}(x)= \frac{\alpha}{2}e^{-\alpha |x|}$ for all $x$.    

\begin{figure}
\begin{center}
    \includegraphics[width=10cm]{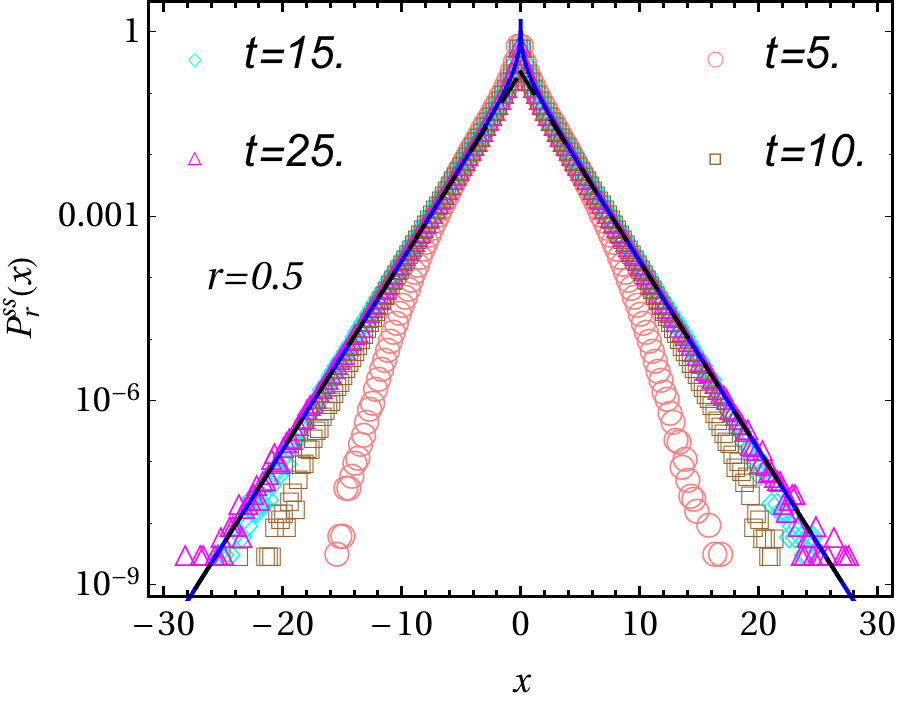}{\centering}
\end{center}
   \caption{\label{prob-1} A comparison of the analytical result of the probability density function $P_r^{ss}(x)$ with respect to the position $x$ with the numerical simulation is shown. The numerical simulation results are shown at four different times: $t=5$ (pink circle), $t=10$ (brown square), $t=15$ (cyan rhombus ), and $t=25$ (magenta triangle). We compare the analytical steady state distribution $P_r^{ss}(x)$ (blue solid line) given in \eref{ss-reset} with the numerical simulations. Moreover, the large-$x$ behaviour of the steady state distribution (black dashed line) given in \eref{asym} is also plotted and has good agreement with both numerical simulations and analytical probability density function (blue solid line). All these results are shown for $t_\gamma=1$, $D=1$, $r=0.5$, and time step $\Delta t=10^{-3}$. The plot indicates that as the time increases, the position distribution reaches steady state and the tails of distribution follow \eref{asym}. }
\end{figure}

In \fref{prob-1}, we present the numerical simulation results for the probability density function for the position of the Brownian particle at four different times : $t=5$ (pink circle), $t=10$ (brown square), $t=15$ (cyan rhombus ), and $t=25$ (magenta triangle). We compare the analytical steady state distribution (blue solid line) given in \eref{ss-reset} with the numerical simulations. Moreover, we plot the asymptotic behaviour of $P_r^{ss}(x)$ (black dashed line) given in \eref{asym}, and there is good agreement between theoretical prediction and the numerical simulation results at large time. These results are shown for $\tau_\gamma=1$, $D=1$, $r=0.5$, and time step $\Delta t=10^{-3}$.

\section{Relaxation to the nonequilibrium steady state}
\label{sec-rlx}
In the previous section, we compare the analytical result of the nonequilibrium steady state of the position distribution with the numerical simulations. It can be seen that the agreement between those two results is good up to a certain range of $x$ around the resetting location $x=0$, and after that analytical result deviates from their numerical counterparts as expected. To understand this, we compute $P_r(x,t)$ given in \eref{renewal-eqn} for a finite time. 

\begin{figure}
\begin{center}
    \includegraphics[width=10cm]{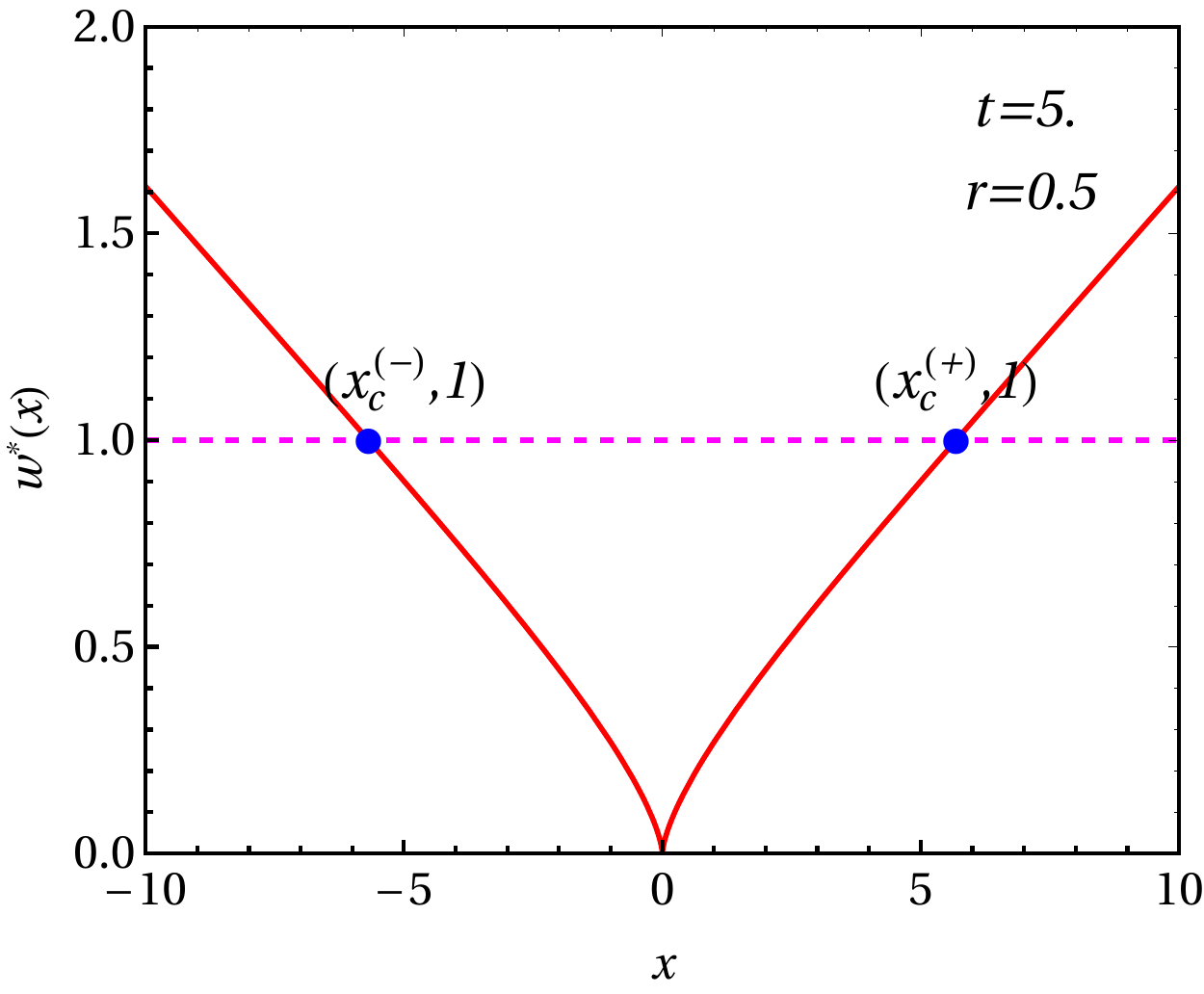}
\end{center}
   \caption{\label{SP} The saddle-point $w^*(x)$ as a function of $x$ is shown for fixed parameters $D=1$, $\tau_\gamma=1$, $r=0.5$, and $t=5$. Blue points denote the point of intersections of $w^*(x)$ and $w^*=1$. }
\end{figure}

\begin{figure}
\begin{center}
    \includegraphics[width=7cm]{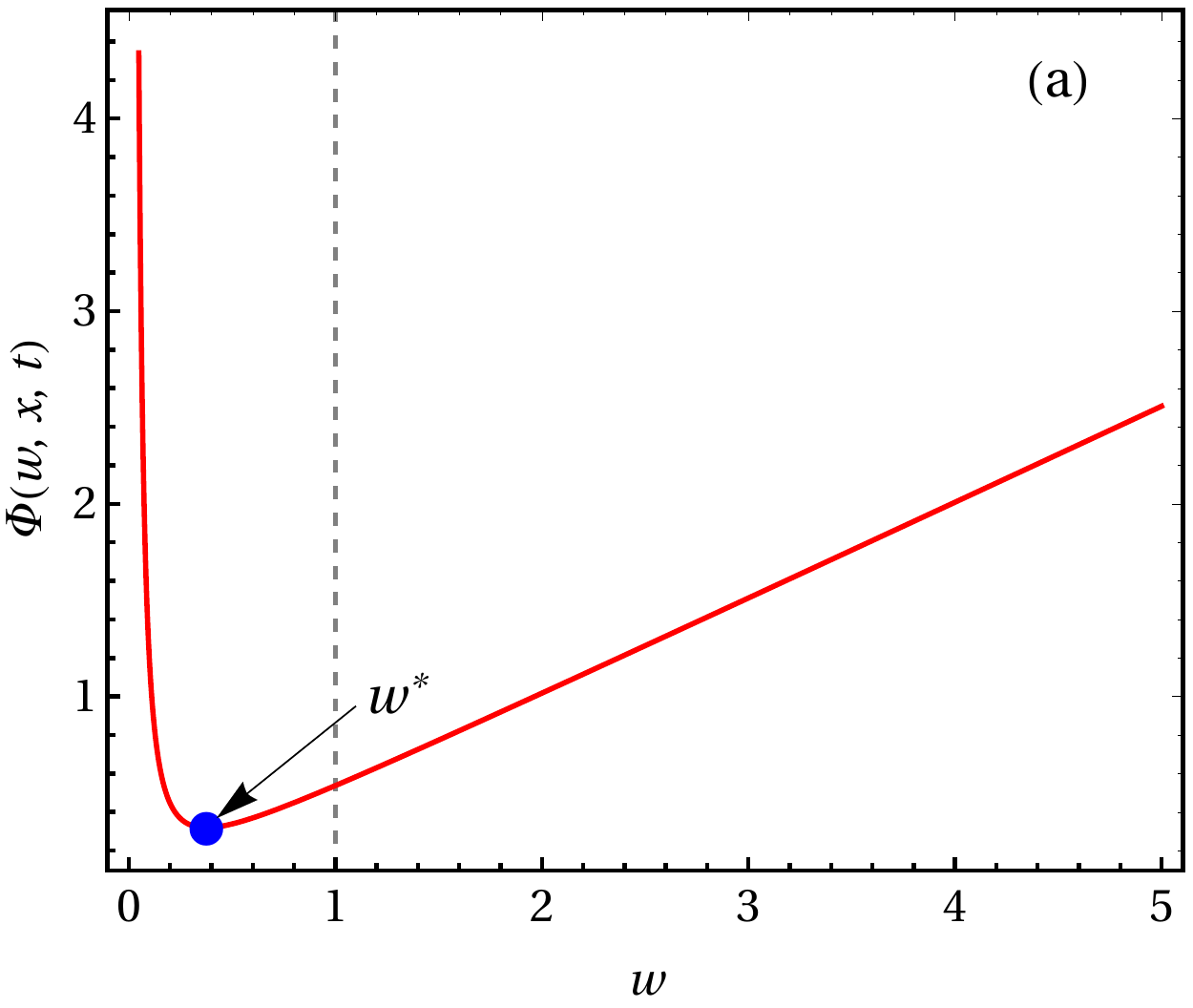} ~~~~
    \includegraphics[width=7cm]{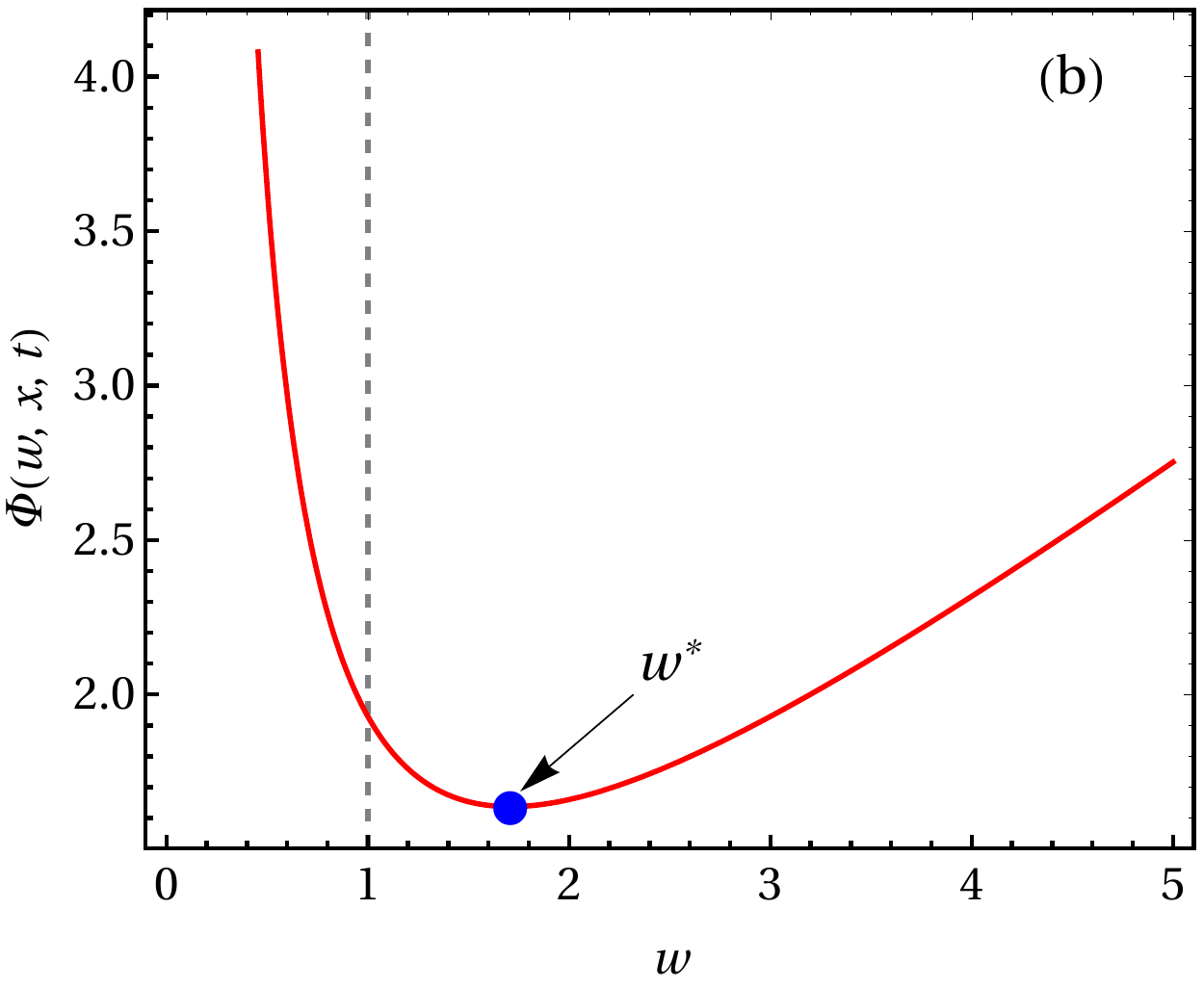}
\end{center}
   \caption{\label{sit} Two situations are shown at fixed $t$ and different values of $x$: (a) the minimum of the function $\Phi(w,x,t)$, i.e., $w^*$, lies inside the limit of the integration [0,1] [see \eref{renewal}] and (b) $w^*>1$. The vertical dashed lines are drawn at $w=1$.}
\end{figure}

Using $P_0^{\mathrm{(I)}}(x,t)$ and $P_0^{\mathrm{(II)}}(x,\tau)$ given in \erefss{prop-1} and \erefs{prop-2}, respectively, a change of variable $w=\tau/t$, and  following \cite{sanjib-relaxation}, we rewrite the equation \eref{renewal-eqn} as 
\begin{align}
P_r(x,t)=\dfrac{e^{-t\Phi(1,x,t)}}{\sqrt{4 \pi D \tau_\gamma g(1,t)}}+\dfrac{rt}{\sqrt{4 \pi D \tau_\gamma}}\int_0^1 dw\ \dfrac{e^{-t\Phi(w,x,t)}}{\sqrt{g(w,t)}},
\label{renewal}
\end{align} 
where 
\begin{align}
&g(w,t)=wt/\tau_\gamma+e^{-wt/\tau_\gamma}-1, \quad\quad \text{and}\\
&\Phi(w,x,t)=r w+\dfrac{x^2}{4 D\tau_\gamma t\ g(w,t)}.
\end{align}
For given $x$ and at large time $t$, one can approximate the integral given in the second term using the saddle-point method. Therefore, the equation
\begin{align} 
\dfrac{\partial \Phi(w,x,t)}{\partial w}\bigg|_{w=w^*}=0
\end{align}
gives the saddle-point $w^*(x)$. It is difficult to obtain the analytical expression for the saddle-point $w^*$. Nevertheless, one can compute it numerically. For a given parameters $t$, $\tau_\gamma$, $r$, $D$, the saddle-point $w^*(x)$ moves as $x$ increases from $-\infty$ to $+\infty$ (see \fref{SP}). Depending on the values of $x$, two situations arise: (1) either $w^*<1$ or (2) $w^*>1$. In the first situation, the function $\Phi(w,x,t)$ has a minimum $w^*$ within the integration limits, i.e., $w^*\in [0,1]$ (see \fref{sit}a). Therefore, the dominant contribution to the above given integral comes from the saddle-point approximation, i.e., $P_r(x,t)\sim e^{-t \Phi(w^*,x,t)}$. In the second situation, the function $\Phi(w,x,t)$ has a lowest value at the upper limit of the integral (see \fref{sit}a), i.e., at $w=1$. Therefore, the integral can be estimated at that limit, i.e., $P_r(x,t)\sim e^{-t \Phi(1,x,t)}$, and has the same order as the first term in \eref{renewal}.  

In \fref{SP}, we plot saddle-point $w^*(x)$ with respect to $x$ for fixed parameters $D=1$, $\tau_\gamma=1$, $r=0.5$, and $t=5$. The curve $w^*(x)$ intersects with the line at $w^*=1$ at two points $x_c^{(\pm)}$ on the $x$-axis  (see blue points). These points are the solutions of the equation $\Phi'(w^*=1,x_c^{(\pm)},t)=0$, where $\prime$ indicates the derivative with respect to $w$. Therefore, we get
\begin{align}
x_c^{(\pm)}=\pm\sqrt{\dfrac{4 r D \tau_\gamma^2}{1-e^{-t/\tau_\gamma}}}g(1,t).
\label{xc}
\end{align}
Notice that in the overdamped limit ($\tau_\gamma\to 0$), $x_c^{(\pm)}\propto t$ as observed in Ref. \cite{sanjib-relaxation}.

Therefore, the probability density function has the following behaviour (ignoring the prefactor)
\begin{align}
P_r(x,t)\sim\begin{cases} e^{-t\Phi(w^*,x,t)} \quad\quad&\text{for} \quad |x|<|x_c^{(\pm)}|,\\
e^{-t\Phi(1,x,t)} \quad\quad &\text{for} \quad |x|>|x_c^{(\pm)}|.
\end{cases}
\end{align}

The complete probability density function $P_r(x,t)$ for the position $x$ at large time $t$ is given by 
\begin{align}
P_r(x,t)=\mathcal{A}\bigg[\dfrac{r t e^{-t \Phi(w^*,x,t)}R_1(x,t)}{\sqrt{4 \pi D\tau_\gamma g(w^*,t)}}\Theta\big(|x_c^{(\pm)}|-|x|\big)+\dfrac{r t e^{-t \Phi(1,x,t)}R_2(x,t)}{\sqrt{4 \pi D\tau_\gamma g(1,t)}}\Theta\big(|x|-|x_c^{(\pm)}|\big)\bigg],
\label{full-pdf}
\end{align} 
where $\Theta(y)$ is the Heaviside-Theta function, and $\mathcal{A}$ is the normalization constant such that $\int_{-\infty}^{+\infty}P_r(x,t)=1$.
In the above equation, 
\begin{align}
&R_1(x,t)=\sqrt{\dfrac{\pi}{2 a}}\bigg[\mathrm{erf}\bigg(b\sqrt{\frac{a}{2}}\bigg)-\mathrm{erf}\bigg((b-1)\sqrt{\frac{a}{2}}\bigg)\bigg],\\
&R_2(x,t)=e^{\frac{c^2}{2d}}\sqrt{\dfrac{\pi}{2 d}}\bigg[\mathrm{erf}\bigg(\frac{c}{\sqrt{2d}}\bigg)-\mathrm{erf}\bigg(\frac{c-d}{\sqrt{2d}}\bigg)\bigg],
\end{align}
where $a=t \Phi''(w^*,x,t)$, $b=w^*$, $c=t\Phi'(1,x,t)$, $d=t\Phi''(1,x,t)$, and erf(u) is the error function defined as $\mathrm{erf}(u)=\frac{2}{\sqrt{\pi}}\int_0^u dz\ e^{-z^2}.$
\begin{figure}
\begin{center}
    \includegraphics[width=10cm]{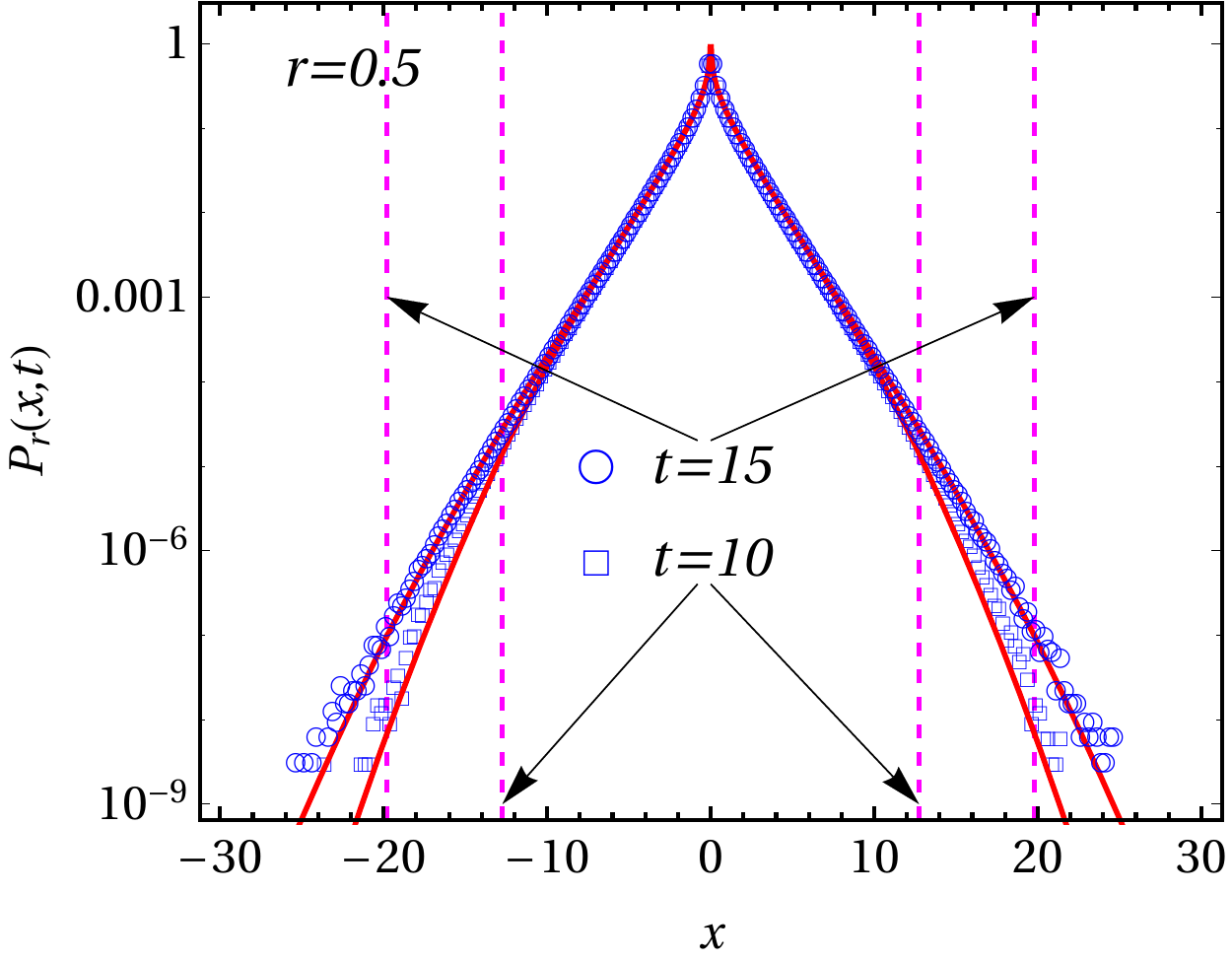}{\centering}
\end{center}
   \caption{\label{prob-2} The analytical result (red solid lines) of the probability density function [see \eref{full-pdf}] of the position of a Brownian particle following dynamics \erefss{pos-1} and \erefs{vel-1} and being stochastically reset to the origin of the $x$-axis with a constant rate $r$ is compared with the numerical simulations. We show the comparison for two different times: $t=10$ (blue squares), and $t=15$ (blue circles). The vertical magenta dashed lines correspond to $x_c^{(\pm)}$ given in \eref{xc} at respective times. The plot is shown for $D=1$, $\tau_\gamma=1$, $r=0.5$, and $\Delta t=10^{-3}$. }
\end{figure}

\ffref{prob-2} shows the comparison of the analytical result (red solid lines) with the numerical simulations at two different times: $t=10$ (blue squares) and $t=15$ (blue circles). In the figure, the vertical dashed lines correspond to  $x_c^{(\pm)}$ given in \eref{xc} at respective times. The comparison we have shown is for $D=1$, $\tau_\gamma=1$, $r=0.5$, and $\Delta t=10^{-3}$.

From \fref{prob-1}, we see that the analytical probability density function for the position $x$ in the steady state has surprisingly good agreement with the finite time numerical simulation results inside a domain (for example, see blue solid line and cyan rhombus) whereas deviations start appearing outside the domain as expected. This is because the region outside the domain is still transient and the contributions to that region are due to those trajectories which do not reset up to time $t$ at all. The boundaries of the domain within which the theoretical and numerical simulation results agree with each other are given by $x_c^{(\pm)}$. Similar result is also shown in Ref. \cite{sanjib-relaxation}, where the boundaries of the domain move ballistically  as $x_c^{(\pm)}\sim t$. However, in our case, the boundaries of the domain has the form given in \eref{xc}, and in the large time limit $(t\gg \tau_\gamma)$, one observes $x_c^{(\pm)}\sim t$. Therefore, in the large time limit, the probability density function given in \eref{ss-reset} and \eref{full-pdf}  may converge to each other.

\section{Summary}
\label{summ}
We have considered an underdamped Brownian particle of mass $m$ in a heat bath. The state of the system is  described by $(x,v)$ at a time $t$.
The position $x$ of the particle is stochastically reset to the origin of the $x$-axis with a constant rate $r$ while the velocity $v$ of the particle evolves irrespective of the location of the Brownian particle and the resetting mechanism. We have studied the nonequilibrium steady state $P_r^{ss}(x)$ of the position of the Brownian particle using the renewal process. It is found that the steady state distribution has exponential tails, i.e., $P_r^{ss}(x) \to \frac{\alpha}{2}e^{-\alpha^2D\tau_\gamma}e^{-\alpha |x|}\quad \text{as}\quad |x|\to\infty$. The analytical solution of the steady state distribution has good agreement with the  (large time) numerical simulation results inside a domain characterizes by $x_c^{(\pm)}$ whereas outside the domain, it deviates from the numerical counterpart. Therefore, we have studied the finite time probability density function $P_r(x,t)$ for the position variable. We have obtained analytically the probability density function both inside ($|x|<|x_c^{(\pm)}|$) and outside of the domain ($|x|>|x_c^{(\pm)}|$). It is seen that the probability density function inside the given domain reaches a nonequilibrium steady state given by $P_r^{ss}(x)$ whereas it has transient behaviour in the outside the domain as expected. We have also compared the analytical predictions with the numerical simulations, and they have nice agreements.

In the absence of resetting, the joint probability density function for position and velocity in the presence of the absorbing boundary for underdamped Brownian motion and distribution of first passage time have been studied quite extensively \cite{Duck,Marshall,Kainz,Burschka1981}. It would be interesting to understand the statistics of first passage time in the presence of the resetting mechanism in our model system.

Fluctuation theorem for the stochastic observable such as  heat, work, entropy production, etc., is a remarkable result in the nonequilibrium statistical physics and has been tested both theoretically and experimentally. The presence of stochastic resetting violates micro-reversibility. Within this paradigm, Pal et al. \cite{IFT-arnab} studied the Hatano-Sasa relation and integral fluctuation theorem. It would be interesting to study the nature of fluctuation theorem in the presence of stochastic resetting.

\section*{Acknowledgement}
The author thanks Sanjib Sabhapandit and Urna Basu for useful discussions.

\vskip 2cm

\bibliographystyle{unsrt}
\bibliography{reset.bib}
\end{document}